\documentclass[aps,pra,amsfonts,amssymb,amsmath,showpacs,12pt]{revtex4}
\usepackage{graphicx}
\def\ket#1{|#1\rangle}

\def\bracketii#1#2#3{\langle #1|#2|#3 \rangle}
\def\ketbra#1#2{| #1 \rangle \langle #2 |}
\begin{document}
\title{Efficiencies for the single mode operation of a quantum optical nonlinear shift gate}
\author{Kunihiro Kojima$^{1}$}
\email{kuni@es.hokudai.ac.jp}
\author{Holger F. Hofmann$^{1,2}$}
\email{hofmann@es.hokudai.ac.jp}
\author{Shigeki Takeuchi$^{1,2}$}
\email{takeuchi@es.hokudai.ac.jp}
\author{Keiji Sasaki$^{1}$}
\email{sasaki@es.hokudai.ac.jp}
\affiliation{$^{1}$Research Institute for Electronic Science, Hokkaido University, Kita-12 Nishi-6, Kita-ku, Sapporo 060-0812, Japan \\ $^{2}$PRESTO, Japan Science and Technology Corporation (JST), Hokkaido University, Kita-12 Nishi-6, Kita-ku, Sapporo 060-0812, Japan}
\begin{abstract}
We investigate the single mode operation of a quantum optical nonlinear $\pi$ phase shift gate implemented by a single two-level atom in one-dimensional free space. Since the single mode property of the input photons at the atom is not preserved in the interaction at the atom, we analyze the effeciency of single mode operation that can still be achieved. We show how the input pulse shape can be optimized to obtain high efficiencies for the nonlinear single mode operation. With this analysis, we obtain an optimal single mode transmittance per photon of $78$\% for the successful nonliner $\pi$ phase shift operation.
\end{abstract}
%
\maketitle
\section{Introduction}
The progress of photon manipulation technologies can open the door to the implementation of quantum information technologies which may enable us to greatly improve the acquisition, transmittance, and processing of information \cite{nielsen0}.
Since standard optical technologies allow us to control interference effects even at the single photon level, the photon is a strong candidate for quantum information media. In this case, the processing of the binary information encoded into photonic states can be achieved by combining linear optical elements with nonlinear photon-photon switching devices. In particular, a nonlinear phase shift of $\pi$ per photon would be useful to perform conditional operations between a control qubit and a target qubit e.g. in a quantum C-NOT gate \cite{fredkin,nielsen0}. However it is still difficult to achieve such a strong nonlinearity at the single photon level. Single-atom cavity quantum electro dynamics (cavity QED) may offer a possible solution to this problem. This is because a very effective coupling between a single atom and the light field can be achieved in small optical cavities. In fact, an atomic nonlinearity at the single photon level has already been demonstrated experimentally using single atom cavity QED methods \cite{turchette}. Based on a semiclassical analysis, we have pointed out that this method could be used to realize a $\pi$ phase-flip operation at the single photon level if a one-sided cavity is used \cite{holger2}.
However, a semiclassical theory is not sufficient to evaluate the effects of a single atom nonlinearity on arbitrary quantum states of the light field. For this reason, it is necessary to develop fully quantum mechanical descriptions of the interaction between the light field continuum and a single atom \cite{drobny,domokos}.

In our previous work, we have investigated the photon-photon interaction at a two level atom by solving the Schr\"odinger equation of one-dimensional light field propagation to and from the atom \cite{kojima,holger3}.  The results of these investigations have shown that the spatiotemporal coherence of a light field pulse is changed significantly by the interaction with the two level atom. That is, even though the interaction is restricted to a single transversal mode, a single mode model is not sufficient to describe the effects of the single atom non-linearity on an input field, because there is an infinite continuum of longitudinal mode into which the photons can be scattered. However, the encoding of quantum information in optical pulses usually requires that the pulse shape is a well-defined superposition of the longitudinal mode and does not change as a result of the operations performed. It is therefore an open question how much the unavoidable multi-mode scattering will reduce the efficiency of a quantum non-linear shift operation implemented by a single atom in a cavity.

In this paper, we analyze the single mode efficiency of the non-linear shift gate implemented by a single two level atom in a cavity by defining a coherent Gaussian pulse as the single optical mode carrying the quantum information \cite{note}. We can then evaluate the multi-mode output of the field-atom interaction by identifying the components of the output state where all photons are found in the Gaussian pulse mode thus defined. Output photons that have been scattered into other modes by the interaction are then treated as losses. By selecting an appropriate pulse shape and by taking into account a delay time caused by the absorption of photons at the atom, we found that a single mode transmittance of 78\% is possible for the successful implementation of a quantum optical non-linear shift operation.
\section{Response of the atomic nonlinearity to a single mode input pulse}
\begin{figure}[htbp]
\begin{picture}(0,0)
\put(-10,130){(a)}
\put(200,130){(b)}
\end{picture}
\includegraphics[width=7cm]{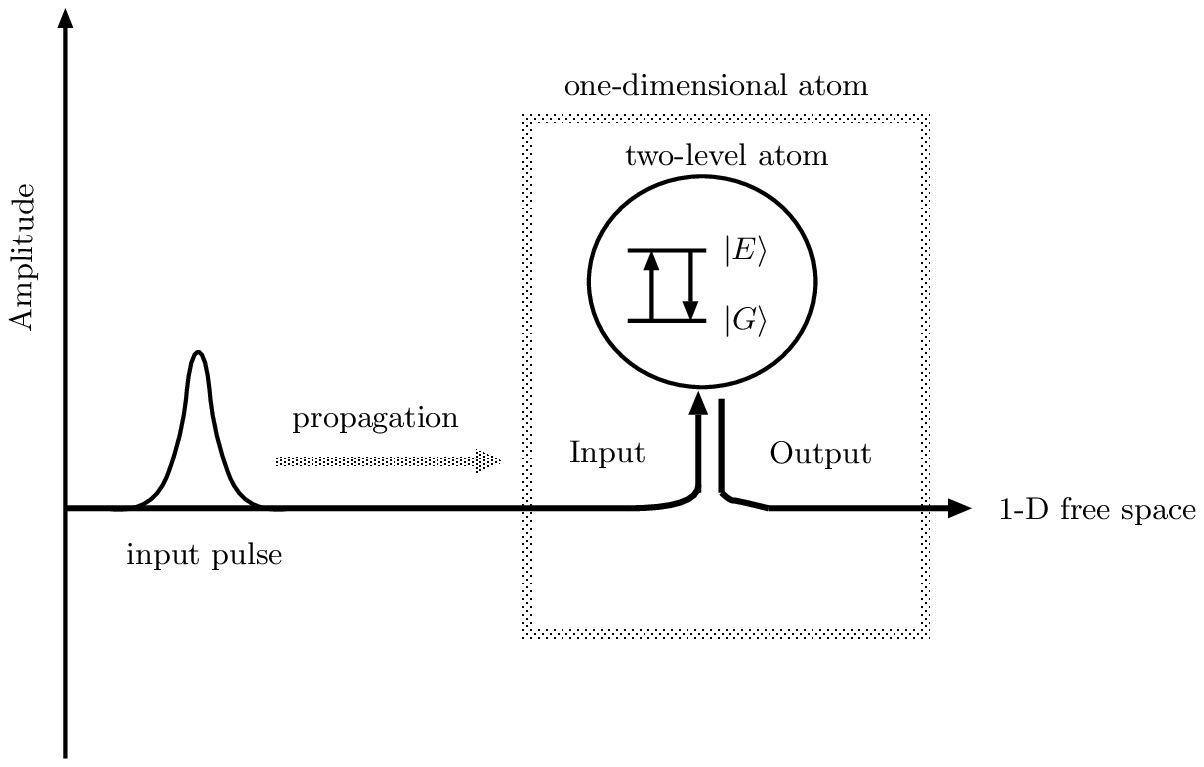}\ \ 
\includegraphics[width=5cm]{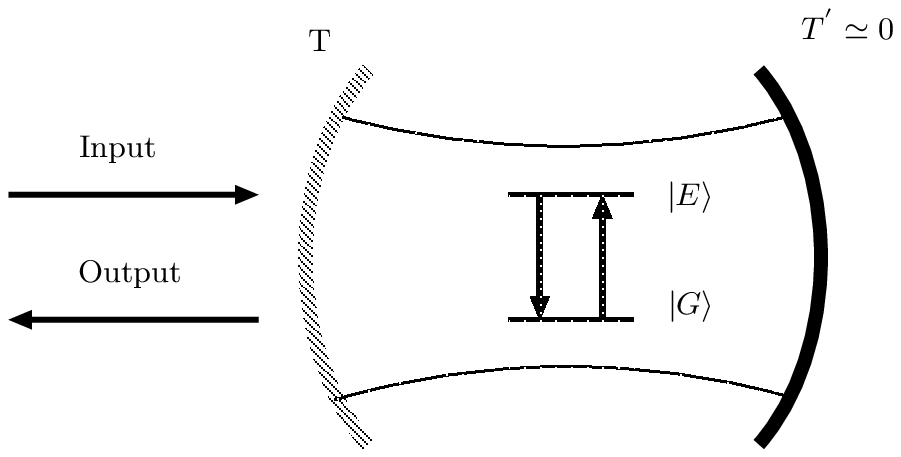}
\caption{\label{fig:ansscheme}(a) Schematic of the atomic nonlinear shift gate and (b) physical realization of the one-dimensional atom.}
\end{figure}

The model of a single two-level atom in one-dimensional free space is shown in Fig.~\ref{fig:ansscheme}(a). The physical realization of this model can be implemented by a two-level atom coupled with a single mode of a one-sided cavity in the bad cavity regime \cite{holger2,kojima}. Here, we assume that the spontaneous emission rate through the cavity mode is much larger than the spontaneous emission rate through non-cavity modes. The cavity geometry is shown in Fig.~\ref{fig:ansscheme} (b). The input field of the one-dimensional free space shown in Fig.~\ref{fig:ansscheme} (a) corresponds to the input of the one-sided cavity in Fig.~\ref{fig:ansscheme} (b), and the output field in Fig.~\ref{fig:ansscheme} (a) corresponds to the output of the one-sided cavity in Fig.~\ref{fig:ansscheme} (b).
Note that we assume perfect mode matching between the input field and the cavity. Likewise, the beam profile of the output field is determined by the emission pattern of the cavity mode. We thus assume an optimized alignment of the linear optics in the setup in order to investigate the inherent limitations of the efficiency caused by the unavoidable changes to the spatiotemporal coherence of the input pulse. In order to apply the non-linearity of the atom-cavity system to an arbitrary quantum state of a single mode light field, we now consider the case of a well-defined input pulse, as indicated in Fig.~\ref{fig:ansscheme} (a). It is then possible to treat the quantum state of this pulse as a single mode quantum state, even though it is actually propagating in the infinite continuum of free space modes.

In order to describe the light field pulse propagating in the free field as a single optical mode, we have to define the pulse shape in terms of the continuous modes of the free space field. Conventionally, this field continuum is represented in terms of the plane wave modes defined by the wave vector $k$. However, since our previous results have been derived in real space, it is more convenient to use the real space representation of the field continuum here. It is then possible to define the creation operator $\hat{\text{A}}_{l}^{\dagger}$ of the single pulse mode as
\begin{eqnarray}
\hat{\text{A}}_{l}^{\dagger} &=& \int \ dx \Psi_{\text A} (x+l) \hat{a}^{\dagger}(x),
\end{eqnarray}
where $\Psi_{\text A} (x+l)$ is the normalized function defining the specific pulse shape. This definition allows us to separate the light field into a single mode A and an infinite number of orthogonal modes that can be distinguished from A. Note that the spatial coordinate $x$ is defined in a coordinate system moving at the speed of light, so that the propagation dynamics of the pulse is effectively included in the definition of the pulse mode \cite{kojima,holger3}. The variable $l$ then represents the precise timing of the pulse. In the following, we will therefore refer to it as the delay of the pulse. As we will show below, this parameter is particularly useful in the analysis of the output. At the moment, it will be sufficient to assume that $l=0$.

Using this creation operator, it is possible to define the spatiotemporal wavefunctions of each photon number state in the pulse. Specifically, the zero, one, and two photon states read
\begin{eqnarray}
\ket{0^{l}_{\text{A}}} &=& \ket{Vac} \nonumber\\
\ket{1^{l}_{\text{A}}} &=& \text{A}_{l}^{\dagger}\ket{Vac} = \int \ dx \Psi_{\text{A}}(x+l) \ket{x} \nonumber\\
\ket{2^{l}_{\text{A}}} &=& \frac{1}{\sqrt{2}}\text{A}_{l}^{\dagger} \text{A}^{\dagger}_{l} \ket{Vac} = \int \ dx_{1}dx_{2} \Psi_{\text{A}}(x_{1}+l) \cdot \Psi_{\text{A}}(x_{2}+l) \ket{x_{1};x_{2}}
\end{eqnarray} 
As explained above, the coordinates $x$, $x_{1}$ and $x_{2}$ are the spatial coordinates in the coordinate system moving at the velocity of light. The indices 1 and 2 of $x_{1}$ and $x_{2}$ in the two photon component identify the two particles. The spatial features of the probability amplitudes given by the wave function $\Psi_{\text{A}}$ correspond to the spatial features of the pulse mode A. Thus, the wave function \(\Psi_{\text{A}}(x)\) characterizes the single mode one photon component. Likewise, the wave function $\Psi_{\text{A}}(x_{1}) \cdot \Psi_{\text{A}}(x_{2})$ characterizes the single mode two photon component, where the wave functions of both particles overlap perfectly. It is thus possible to describe the spatiotemporal feature of the single mode Fock states $\ket{n^{l}_{\text{A}}}$.

Any quantum state of the single mode input pulse can now be expanded in the photon number basis of the pulse mode A. In this basis, an arbitrary input state can be defined by the probability amplitudes $C_{n}$ of the photon number states $\ket{n^{l=0}_{\text{A}}}$.
In the following, we assume that the contributions with photon numbers greater than two are negligible. Note that this situation can be realized either by controlling the photon number precisely, e.g. by using single photon sources, or by using weak coherent input light. In either case, the spatiotemporal coherence of a general single mode input state can then be expressed using the real space representations of the one and two photon components,
\begin{eqnarray}
\ket{\psi}_{\text{in}} &=& C_{0} \ket{0^{l=0}_{\text{A}}} + C_{1} \ket{1^{l=0}_{\text{A}}} + C_{2} \ket{2^{l=0}_{\text{A}}} \nonumber \\
                       &=& C_{0} \ket{Vac} + C_{1} \int \ dx \Psi_{\text{A}}(x) \ket{x} + C_{2} \int \ dx_{1}dx_{2} \Psi_{\text{A}}(x_{1}) \cdot \Psi_{\text{A}}(x_{2}) \ket{x_{1};x_{2}} \label{eq:spatialinput}.
\end{eqnarray}
The light pulse characterized by the above input state is propagating at the velocity of light in the input field of the schematic shown in Fig. \ref{fig:ansscheme}(a).
In our previous work \cite{kojima}, we derived the unitary time evolution operator in the Hilbert spaces for the one photon component and the two photon component by solving the Schr\"odinger equation of the interaction between a one-dimensional field and a single atom. The output state in the far field can then be obtained by integrating the input wave functions with the matrix description of the unitary operator in real space,
\begin{eqnarray}
&& \ket{\psi}_{\text{out}} = C_{0} \ket{Vac} + C_{1} \int \ dx \Psi_{\text{out}}(x) \ket{x} + C_{2} \int \ dx_{1}dx_{2} \Psi_{\text{out}}(x_{1};x_{2}) \ket{x_{1};x_{2}}  \nonumber\\
&& \text{ with } \Psi_{\text{out}}(x) = \int dx^{'} \ {\bf u}_{\text{1photon}} (x;x^{'}) \cdot \Psi_{\text{A}}(x^{'}) \nonumber\\
&& \text{ and }\Psi_{\text{out}}(x_{1};x_{2}) = \int dx^{'}_{1}dx^{'}_{2} \ {\bf u} (x_{1},x_{2};x^{'}_{1},x^{'}_{2}) \cdot \Psi_{\text{A}}(x^{'}_{1}) \cdot \Psi_{\text{A}}(x^{'}_{2}).\nonumber\\
\label{eq:actual}
\end{eqnarray}
The matrix elements ${\bf u} (x_{1},x_{2};x^{'}_{1},x^{'}_{2})$ are given by
\begin{eqnarray} 
&& {\bf u}(x_{1},x_{2};x_{1}^{'},x_{2}^{'}) = {\bf u}_{\text{1photon}}(x_{1};x_{1}^{'}) \cdot {\bf u}_{\text{1photon}}(x_{2};x_{2}^{'}) + \Delta {\bf u}^{{\bf Nonlin}}(x_{1},x_{2};x_{1}^{'},x_{2}^{'}) \nonumber\\
&& \text{ with }{\bf u}_{\text{1photon}}(x;x^{'}) = \delta(x-x^{'})-\frac{2\Gamma}{c} e^{-\frac{\Gamma}{c}(x^{'}-x)} \ \ \ \text{for $x \leq x^{'}$, else $0$} \label{eq:oneoutgene} \nonumber \\
&& \text{ and }\Delta {\bf u}^{{\bf Nonlin}}(x_{1},x_{2};x_{1}^{'},x_{2}^{'}) = -\frac{4 \Gamma^{2}}{c^{2}} e^{-\frac{\Gamma}{c}(x_{1}^{'}+x_{2}^{'}-x_{1}-x_{2})} \ \ \ \text{ for } x_{1},x_{2} < {\bf Min}[x_{1}^{'},x_{2}^{'}]. \nonumber\\
\label{eq:nonlineffect}
\end{eqnarray}
The output wave functions $\Psi_{\text{out}}(x)$ and $\Psi_{\text{out}}(x_{1};x_{2})$ describe the spatial feature of the output state in the far field of the atom shown in Fig.~\ref{fig:ansscheme} (a). The spatiotemporal features of the one photon input and two photon input components are generally not preserved in the output due to the atomic response described by the matrix elements ${\bf u} (x_{1},x_{2};x^{'}_{1},x^{'}_{2})$. This means that the single mode Fock states $\ket{n^{l}_{\text{A}}}$ will not be sufficient to describe the output state, since this state includes photons in other modes. In the following, we therefore analyze which component of the output state can still be described in terms of a single mode pulse state using the single mode photon number basis $\ket{n^{l}_{\text{A}}}$ for a fixed value of the delay $l$.
\section{Single mode output components of the atomic nonlinearity}

In order to analyze what component of the output state can be represented by the single mode basis $\ket{n^{l}_{\text{A}}}$, we have to apply a mathematical operation to the output state that removes any component with photons in modes other than the single pulse mode A. Since this operation effectively filters out any components that cannot be represented as single mode states of the pulse mode A, we will refer to this operation as filter operation in the following. This filter operation can be described by the Hermitian operator
\begin{eqnarray}
\hat{\text{F}}_{\text{A}}(l) &=& \ketbra{Vac}{Vac} + \ketbra{1^{l}_{\text{A}}}{1^{l}_{\text{A}}} + \ketbra{2^{l}_{\text{A}}}{2^{l}_{\text{A}}}.
\end{eqnarray}
Here, $\ket{1^{l}_{\text{A}}}$ and $\ket{2^{l}_{\text{A}}}$ represent the one- and two photon Fock states described by the spatiotemporal wavefunctions $\Psi_{\text{A}}(x)$ and $\Psi_{\text{A}}(x_{1}) \cdot \Psi_{\text{A}}(x_{2})$, respectively. Since the mode selected by the filter operation is a pulse mode, the timing of the filter operation given by the delay $l$ is an important variable. When the delay time is $l=0$, the mode passing the filter corresponds to an input pulse that has passed the atom without absorption and reemission. By increasing the delay time, the delay caused by the temporary absorption of photons can be compensated. Note that the delay time $l$ must be the same for both the one and the two photon components, since the condition for single mode operation is that the two components are Fock states of the same mode.

In order to obtain an intuitive understanding of the above filter process, it may be useful to think of a hypothetical device that could perform the operation described by the operator $\hat{\text{F}}_{\text{A}}(l)$. Such a linear optics filter device would transmit only the pulse mode A with a probability of 1, while all other modes would be reflected. The filter operation would then be controlled by post-selecting only the cases where all photons are transmitted. It should be noted that the practical implementation of such a device may require a significant technological effort, since it must combine time dependent gating with a sufficiently narrow spectral response. In the following, we therefore use the filter operation $\hat{\text{F}}_{\text{A}}(l)$ only as a mathematical tool to identify a well-defined single mode component within the multi-mode output of the non-linear atom-cavity system.

We can now analyze the component of the output state that can be described as a single mode quantum state of a pulse mode A delayed by $l$. If we describe the operation of the atom cavity system by the unitary operator $\hat{\text{u}}_{atom}$ and limit the input to the pulse mode A with $l=0$, the operation of the atom cavity system under the condition that all output photons are in the delayed pulse mode can be described as
\begin{eqnarray}
\hat{\text{S}}_{device} &\equiv& \hat{\text{F}}_{\text{A}}(l)\hat{\text{u}}_{atom} \nonumber\\ 
&=& \ketbra{Vac}{Vac} -  \eta_{1} \ketbra{1^{l}_{\text{A}}}{1^{l=0}_{\text{A}}} - \eta_{2} \ketbra{2^{l}_{\text{A}}}{2^{l=0}_{\text{A}}}, \nonumber
\end{eqnarray}
\begin{eqnarray}
\text{where } \eta_{1} &=& -\bracketii{1^{l}_{\text{A}}}{\hat{\text{u}}_{atom}}{1^{l=0}_{\text{A}}}\nonumber\\ 
                        &=& -\int \ dx \Psi^{*}_{\text{A}}(x+l) \cdot \Psi_{out}(x) \nonumber\\
\ \ \text{and } \eta_{2} &=& -\bracketii{2^{l}_{\text{A}}}{\hat{\text{u}}_{atom}}{2^{l=0}_{\text{A}}} \nonumber\\
 &=& -\int dx_{1}dx_{2}\ \Psi^{*}_{\text{A}}(x_{1}+l)\Psi^{*}_{\text{A}}(x_{2}+l) \cdot \Psi_{out}(x_{1},x_{2}).\nonumber\\
\end{eqnarray}
It is thus possible to identify a conditional single mode operation $\hat{\text{S}}_{device}$ within the non-linear multi-mode operation $\hat{\text{u}}_{atom}$. For the analysis of the single mode efficiency of the atomic device $\hat{\text{S}}_{device}$, it will be useful to separate this operation into two components, a nonlinear phase shift operation $\hat{\text{U}}_{NS}$ and a loss operation $\hat{\text{L}}$,
\begin{eqnarray}
&& {} \hat{\text{S}}_{device} = \hat{\text{L}}\hat{\text{U}}_{NS}, \nonumber\\
&& \text{ where } \hat{\text{L}} = \ketbra{Vac}{Vac} + \eta_{1} \ketbra{1^{l}_{\text{A}}}{1^{l}_{\text{A}}}+\eta_{2} \ketbra{2^{l}_{\text{A}}}{2^{l}_{\text{A}}} \nonumber\\
&& \text{ and } \hat{\text{U}}_{NS} = \ketbra{Vac}{Vac}-\ketbra{1^{l}_{\text{A}}}{1^{l=0}_{\text{A}}}-\ketbra{2^{l}_{\text{A}}}{2^{l=0}_{\text{A}}}. \nonumber\\
\end{eqnarray}
The loss operation then determines the probability of a successful single mode operation,
\begin{eqnarray}
\text{P}_{success} &=& \bracketii{\psi}{\hat{L}^\dagger
\hat{L}}{\psi}_{out}
\nonumber \\
&=& |C_0|^2 + \eta_1^2 |C_1|^2 + \eta_2^2 |C_2|^2.
\end{eqnarray}
The probability of success thus depends on the photon number distribution of the input state. Since the probability of success for the vacuum component is always one, this dependence can be conveniently expressed in terms of a transmittance per photon. The single photon transmittance is then given by $\eta_1^2$, while the transmittance per photon in the two photon case is given by $\eta_2$. As long as the sign of $\eta_{2}$ is positive, the operation $\hat{\text{S}}_{device}$ thus represents a single mode nonlinear $\pi$ phase shift operation $\hat{\text{U}}_{NS}$ with photon losses given by $\hat{\text{L}}$. Note that, if $\eta_{2}$ becomes negative, $\hat{\text{L}}$ itself causes a nonlinear phase shift, canceling the effect of $\hat{\text{U}}_{NS}$.

The single mode transmittance of the non-linear shift operation can now be optimized by varying the pulse shape A and the delay $l$ between the input and the output. It is then possible to determine optimized conditions for the single mode operation on an arbitrary input state.


\section{Optimization of the nonlinear shift gate}
We can now find the optimal effeciency of the nonlinear operation by selecting the delay time $l$ and the input pulse length. In the following, we consider Gaussian input pulses with varying pulse lengths given by
\begin{eqnarray}
&& \Psi_{\text{A}}(x) = e^{-x^{2}/(cT)^{2}}/\sqrt{N}, \nonumber \\
&& \text{ where }N=\sqrt{(\pi/2)(cT)^{2}}.
\end{eqnarray}
The transmittances depend strongly on the input pulse length and the delay time of the filter, since photon absorption and reemission processes are sensitive to the spatiotemporal coherence of the light field.
In general, the loss operator describes nonlinear losses. This means that the transmittance per photon is different in the one photon and the two photon case. To optimize both cases at the same time, we choose the condition of linear transmittance, $\eta_{1}^{2}=\eta_{2}$. This condition allows us to interpret the transmittance as a linear optics effect where each photon has the same probability of being lost. In particular, this condition greatly simplifies the calculation of total losses in complicated networks.
Note that the condition of linear transmittance can always be 
fulfilled by adjusting the delay time $l$ of the output filter 
at a fixed pulse length $T$. The optimization of the 
transmittance can then be accomplished by varying the only 
remaining parameter $T$.

\begin{figure}[htbp]
\begin{picture}(0,0)
\put(-20,115){(a)}
\put(210,115){(b)}
\end{picture}
\includegraphics[width=7cm]{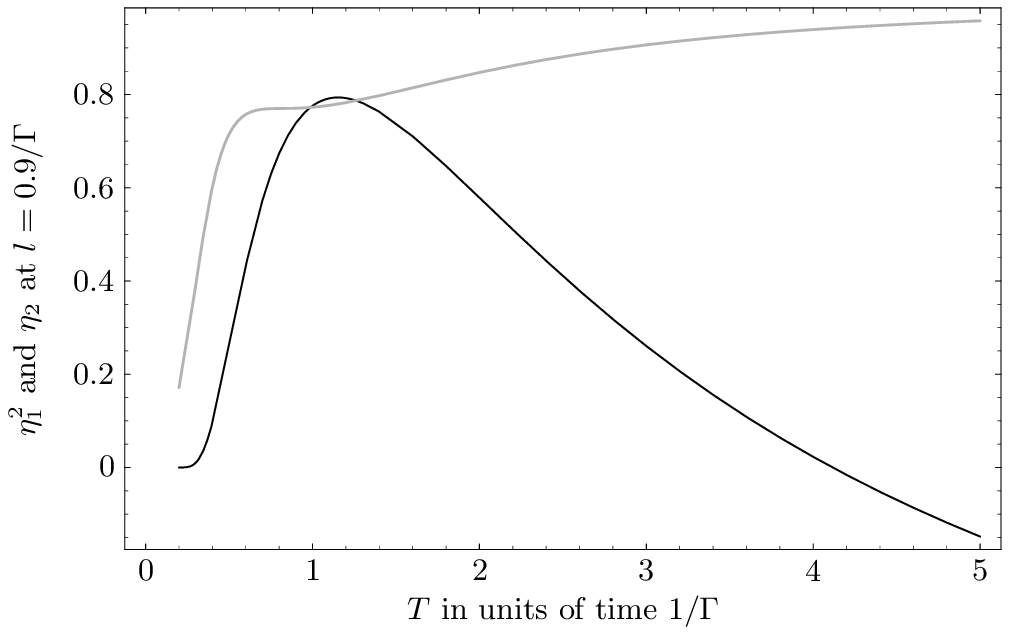}
\includegraphics[width=7cm]{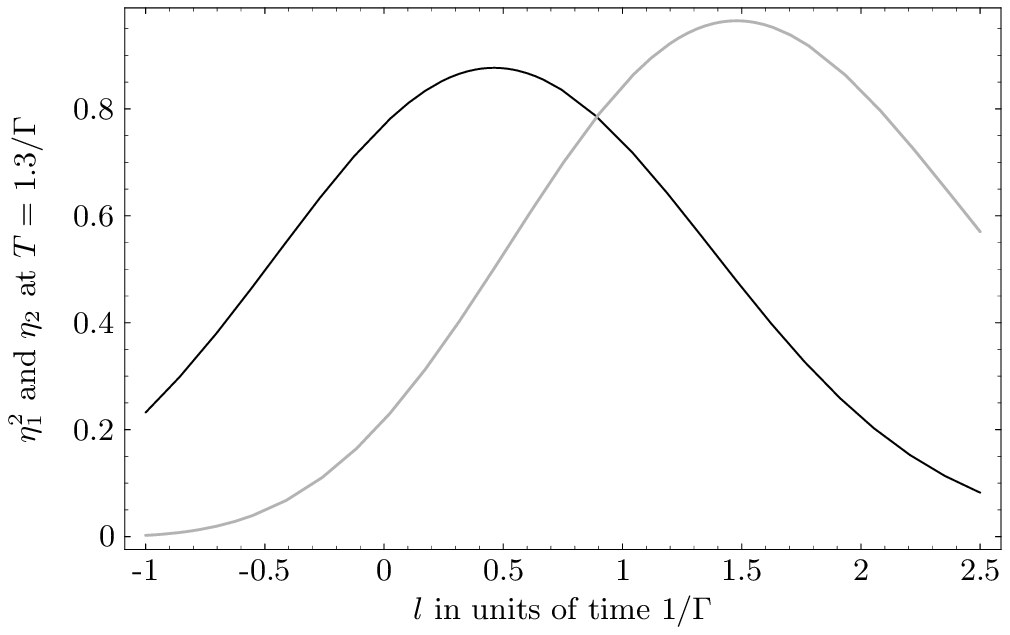}
\caption{\label{fig:nstester} (a) Transmittances $\eta_{1}^{2}$ (thin line) and $\eta_{2}$ (solid line) depending on the pulse duration $T$ at a delay time of $l=0.9/\Gamma$ and (b) transmittances $\eta_{1}^{2}$ (thin line) and $\eta_{2}$ (solid line) depending on the delay time $l$ at a pulse duration of $T=1.3/\Gamma$. The condition for the linear transmittance, $\eta_{2}=\eta_{1}^{2}$, is fulfilled at the crossing points of the two lines.}
\end{figure}

The maximal transmittance of $\eta^{2}_{1}=0.78$ under the condition of linear transmittance $\eta_{1}^{2}=\eta_{2}$ has been obtained numerically in the vicinity of a pulse duration of $T=1.3/\Gamma$ and a delay time of $l=0.9/\Gamma$ by varying the input pulse duration $T$ and the delay time of the filter $l$.
 The dependence of the transmittances $\eta^{2}_{1}$ and $\eta_{2}$ on the pulse duration $T$ at the delay time of $l=0.9/\Gamma$ is shown in Fig.\ref{fig:nstester} (a). The thin line and the solid line correspond to the one photon transmittance $\eta_{1}^{2}$ and the two photon transmittance $\eta_{2}$, respectively. The one photon transmittance $\eta^{2}_{1}$ increases rapidly until up to a pulse duration of $T=0.5/\Gamma$ and then slowly approaches one.
For input pulse durations much smaller than the dipole relaxation time $1/\Gamma$, the input pulse suffers almost no delay by absorption and reemission. The low one photon transmittance $\eta^{2}_{1}$ shown in Fig. \ref{fig:nstester} (a) for such short pulses therefore originates from the difference between the delay time $l=0.9/\Gamma$ of the filter and the actual arrival time of the output pulse corresponding to a delay of $l=0$. For long pulses, the one photon output component becomes almost identical to the one photon input component in terms of pulse length and pulse shape, and the delay in the arrival time is now much smaller than the input pulse length $T$. The one photon transmittance $\eta^{2}_{1}$ therefore approaches one, as shown in Fig. \ref{fig:nstester} (a).
On the other hand, the two photon transmittance $\eta_{2}$ has a maximum near a pulse duration of $T=1/\Gamma$, and then begins to decrease, eventually approaching minus one where the output two photon wave function at the atom is identical with the input two photon wave function except for the delay time $l$. This is because the nonlinear effect only occurs if the two photons are within about $1/\Gamma$ of each other \cite{kojima}. Therefore, the nonlinear response of the atom is maximal near a pulse duration of $T=1/\Gamma$ and the linear response is dominant for input pulses much longer than $T=1/\Gamma$.

The dependence of the transmittances $\eta^{2}_{1}$ and $\eta_{2}$ on the delay time $l$ at the pulse duration of $T=1.3/\Gamma$ is shown in Fig.\ref{fig:nstester} (b). The thin line and the solid line correspond to the one photon transmittance $\eta_{1}^{2}$ and the two photon transmittance $\eta_{2}$, respectively. The maxima of the one photon transmittance and the two photon transmittance are delayed due to absorption at the atom. The difference of the delay time of the two maxima is roughly equal to the dipole relaxation time $1/\Gamma$. Specifically, the one photon maximum is delayed compared to the two photon maximum. This difference between the two cases suggests that the reemission from the atom occurs sooner in the two photon case. It can therefore be interpreted as an effect of stimulated emission \cite{kojima0}. This delay between the one photon and two photon output  appears to be one of the reasons why we cannot get beyond a linear transmittance of $0.78$.

The optimal linear transmittance is given by the crossing points of the one photon and two photon transmittances in Fig.\ref{fig:nstester} (a) and Fig.\ref{fig:nstester} (b). Note that, in Fig.\ref{fig:nstester} (a), there are two crossing points at a pulse duration of $T=1.0/\Gamma$ and a pulse duration of $T=1.3/\Gamma$. The change of transmittance in the region between the two points is only about 0.01. This means that the linear transmittances is not very sensitive to the input pulse duration between $T=1.0/\Gamma$ and $T=1.3/\Gamma$, which may make it easier to optimize pulse duration in experiments. The result of $0.78$ at a pulse duration of $T=1.3/\Gamma$ and a delay time of $l=0.9/\Gamma$ means that a successful nonlinear $\pi$ phase shift can be achieved where $78$\% of the input photons are transmitted in single mode by the nonlinear device. 
Our result shows how the choice of the pulse length and the output delay time affects the nonlinear operation of the single
atom device. The efficiency of $78$\% is the maximal efficiency achieved with Gaussian input pulses when there are no additional losses in the device. In the presence of losses, induced e.g. by imperfect mode matching of the input beam and the cavity mode or by spontaneous emission into non-cavity modes, the efficiency will be reduced accordingly. While a detailed analysis of such losses is beyond the scope of this paper, it may be worth noting that linear losses could be included in the model by applying an additional linear loss operator. A simple estimate of the total efficiency can then be obtained by multiplying the idealized efficiency with the actual linear transmittivity of the device at resonance.  
\section{Conclusions}
In conclusion, we 
have
investigated the effeciency for the single mode operation of a nonlinear $\pi$ phase shift implemented using an atom-cavity system by analyzing the multi mode property of the output photons from the one-dimensional atom. The maximal transmittance of $\eta^{2}_{1}=0.78$ under the condition of linear transmittance $\eta_{1}^{2}=\eta_{2}$ has been obtained numerically in the vicinity of a pulse duration of $T=1.3/\Gamma$ and a delay time of $l=0.9/\Gamma$ by varying the input pulse duration $T$ and the delay time of the filter $l$. The numerical results for Gaussian input pulses indicates that the effeciency is very sensitive to the input pulse length and to the delay time of the pulse mode filter. Our analysis for the single mode operation clearly shows the importance of the input pulse length and the delay time for the realization of successful single mode operation using atomic devices. Specifically, the optimized effeciency of $78$\% for single mode operation indicates that nonlinear devices based on atomic nonlinearities could indeed be useful for quantum information processing if proper pulse lengths and delay lines are used.
\begin{acknowledgments}
This work was partly supported by the program "Research and Development on Quantum Communication Technology" of the Ministry of Public Management, Home Affairs, Posts and Telecommunications of Japan.
\end{acknowledgments}


\begin{thebibliography}{99}
\bibitem{note}Note that we can neglect other levels of the atom, since two levels are sufficient for the self-phase modulation effect considered here.
\bibitem{kojima0}Note that the stimulated emission can also be interpreted as an effect of quantum interference of different processes in the light-atom interaction as discussed in \cite{kojima}.
\end{thebibliography}
\end{document}